\documentclass[twocolumn]{aastex61}
\usepackage{graphicx}

\newcommand\aastex{AAS\TeX}

\shorttitle{\aastex\ LHS~1610A}
\shortauthors{Winters et al.}

\begin{document}

\title{LHS~1610A: A Nearby Mid-M Dwarf with a Companion That is Likely
  A Brown Dwarf }

\correspondingauthor{Jennifer G. Winters}
\email{jennifer.winters@cfa.harvard.edu}

\author{Jennifer G. Winters}
\affil{Harvard-Smithsonian Center for Astrophysics \\
60 Garden Street \\
Cambridge, MA 02138, USA}

\author{Jonathan Irwin}
\affiliation{Harvard-Smithsonian Center for Astrophysics \\
60 Garden Street \\
Cambridge, MA 02138, USA}

\author{Elisabeth R. Newton}
\affiliation{Massachusetts Institute of Technology \\
MIT Kavli Institute for Astrophysics and Space Research \\
77 Massachusetts Ave., Building 37-675 \\
Cambridge, MA 02109, USA}

\author{David Charbonneau}
\affiliation{Harvard-Smithsonian Center for Astrophysics \\
60 Garden Street \\
Cambridge, MA 02138, USA}

\author{David W. Latham}
\affiliation{Harvard-Smithsonian Center for Astrophysics \\
60 Garden Street \\
Cambridge, MA 02138, USA}

%

\author{Eunkyu Han}
\affiliation{Department of Astronomy \& The Institute for Astrophysical Research\\
Boston University \\
725 Commonwealth Avenue \\
Boston, MA 02215, USA}

\author{Philip S. Muirhead}
\affiliation{Department of Astronomy \& The Institute for Astrophysical Research\\
Boston University \\
725 Commonwealth Avenue \\
Boston, MA 02215, USA}

\author{Perry Berlind}
\affiliation{Harvard-Smithsonian Center for Astrophysics \\
60 Garden Street \\
Cambridge, MA 02138, USA}

\author{Michael L. Calkins}
\affiliation{Harvard-Smithsonian Center for Astrophysics \\
60 Garden Street \\
Cambridge, MA 02138, USA}

\author{Gil Esquerdo}
\affiliation{Harvard-Smithsonian Center for Astrophysics \\
60 Garden Street \\
Cambridge, MA 02138, USA}

\begin{abstract}

We present the spectroscopic orbit of LHS~1610A, a newly discovered
single-lined spectroscopic binary with a trigonometric distance
placing it at 9.9 $\pm$ 0.2 pc. We obtained spectra with the TRES
instrument on the 1.5m Tillinghast Reflector at the Fred Lawrence
Whipple Observatory located on Mt. Hopkins in AZ. We demonstrate the
use of the TiO molecular bands at 7065 -- 7165~\AA ~to measure radial
velocites and achieve an average estimated velocity uncertainty of 28
m s$^{-1}$. We measure the orbital period to be 10.6 days and
calculate a minimum mass of $44.8 \pm 3.2$ M$_{\rm Jup}$ for the
secondary, indicating that it is likely a brown dwarf. We place an
upper limit to 3$\sigma$ of 2500 K on the effective temperature of the
companion from infrared spectroscopic observations using IGRINS on the
4.3m Discovery Channel Telescope. In addition, we present a new
photometric rotation period of 84.3 days for the primary star using
data from the MEarth-South Observatory, with which we show that the
system does not eclipse.

\end{abstract}


\section{Introduction} \label{sec:intro}

The nearest stars provide the best representatives of their kinds for
study, with the canonical 10 pc sample containing the most easily
targeted sample of stars. Remarkably, discoveries within this volume
continue to be made, especially amongst the M dwarf population. New
members of note include both M dwarf primaries and their stellar and
sub-stellar companions, as reported in
\citet{Deacon(2005a),Biller(2006),Henry(2006),Winters(2011),Davison(2014)}.

In number, M dwarfs make up 75\% of all stars \citep{Henry(2006)}, but
have historically been challenging targets to study due to their low
luminosities. This has been especially true in the field of high
resolution spectroscopy, which typically requires bright
targets. Thus, many faint, nearby M dwarfs lack high-resolution
spectroscopic measurements. However, the combination of modern
echelle/CCD spectrographs with new analysis techniques allow this
population of stars to benefit from higher resolution instrumentation.


Multiplicity studies contribute to a better understanding of star and
planet formation, as the shape of mass ratio distributions provides
hints as to which pairs of stars are preferentially formed. Equal-mass
(and therefore, equal-luminosity) companions are typically the most
easily studied. Low-mass companions contribute very little light to
the system and are therefore more challenging to detect. Companions
that are both low-mass {\it and} members of short orbital period
binaries can usually be detected only via the radial velocity method,
as their corresponding angular separations are too small to resolve
with other techniques such as astrometry, adaptive optics imaging,
lucky imaging or speckle interferometry. Because the mass ratio
distribution for M dwarfs is not yet well measured at small mass
ratios (where mass ratio $q =$ $M_{\rm sec}/M_{\rm pri}$ $<$ 0.50 and
where $M_{\rm pri}$ and $M_{\rm sec}$ represent the masses of the
primary and secondary components, respectively), the identification
and characterization of short-period low-mass companions, in
particular, is critical to understanding the shape of the
distribution. This can only be accomplished with high-resolution
spectroscopic work.

While it has been shown that stellar companions are less common around
M dwarfs than around more massive stars
\citep{Henry(1991),Fischer(1992),Duchene(2013),Janson(2014a),Wintersphd,Ward-Duong(2015)},
brown dwarf companions to M dwarfs are even more rare. Few examples
are known, despite significant efforts to identify them in the solar
neighborhood
\citep{Campbell(1988),Marcy(1989),Henry(1990),Tokovinin(1992b),Dieterich(2012)}.
Only four M dwarf - brown dwarf pairs are known within 10
pc. Additions to this meagre population provide precious data points
to aid in constraining star and planet formation and evolution models.

We are conducting a multi-epoch spectroscopic survey of a
volume-complete all-sky sample of 456 stars with estimated masses 0.1
-- 0.3 M$_{\odot}$ and with trigonometric distances placing them
within 15 pc. During the course of our observations, we discovered a
previously unknown single-lined spectroscopic binary: \object[Wolf
  227]{LHS~1610A}. Here we present the characterization of this
system.




\begin{deluxetable}{lcc}
\tablecaption{System Parameters for LHS~1610A \label{tab:system_info}}
\tablecolumns{3}
\tablenum{1}
\tablewidth{0pt}
\tablehead{
\colhead{Parameter}  &
\colhead{Value}      &
\colhead{Reference}      
}

\startdata
RA (2000.0) (hh:mm:ss)                     & 03:52:41.8          & 2 \\
Decl. (2000.0) (dd:mm:ss)                  & $+$17:01:04         & 2 \\
Proper Motion Mag. (mas yr$^{-1}$)          & 767 $\pm$ 1.0       & 2 \\
Proper Motion PA (deg)                     & 146 $\pm$ 0.15      & 2 \\
Parallax (mas)\tablenotemark{a}            & 100.88 $\pm$ 2.05   & 2,4 \\
$V_{\rm J}$ (mag)                            & 13.79 $\pm$ 0.02    & 5 \\
$R_{\rm KC}$ (mag)                           & 12.42 $\pm$ 0.02    & 5 \\
$I_{\rm KC}$ (mag)                           & 10.67 $\pm$ 0.02    & 5 \\
$J$ (mag)                                  & 8.93 $\pm$ 0.03     & 3 \\
$H$ (mag)                                  & 8.38 $\pm$ 0.03     & 3 \\
$K_{\rm S}$ (mag)                            & 8.05 $\pm$ 0.02     & 3 \\
Primary mass (M$_{\odot}$)\tablenotemark{b}  & 0.17  $\pm$ 0.02   & 1 \\
Spectral Type                              & M4.0 V             & 2 \\
Rotation Period (days)\tablenotemark{c}    & 84.3               & 1 \\
$U_{\odot}$ (km s$^{-1}$)\tablenotemark{d}    & -30.5 $\pm$ 0.4    & 1 \\
$V_{\odot}$ (km s$^{-1}$)\tablenotemark{d}    & -32.0 $\pm$ 0.7    & 1 \\
$W_{\odot}$ (km s$^{-1}$)\tablenotemark{d}    & -21.3 $\pm$ 0.3    & 1 \\
\enddata
\tablenotetext{a}{Weighted mean parallax.}
\tablenotetext{b}{Estimated using the $M_K$~mass-luminosity relation from \citet{Benedict(2016)}.}
\tablenotetext{c}{As reported in \citet{Irwin(2011)}, signal injection
  and recovery tests indicate that uncertainties on MEarth period
  measurements are 5\% -- 10\% for periods between 50 and 100 days.}
\tablenotetext{d}{Space motions relative to the Solar System.}
\tablerefs{
(1) this work;
(2) \citet{Henry(2006)};
(3) \citet{Skrutskie(2006)};
(4) \citet{vanAltena(1995)};
(5) \citet{Weis(1996)}.}
\end{deluxetable}



\section{Data Acquisition} \label{sec:data}

We obtained 13 optical spectra between UT 2017 February 1 and 2017
March 12 using the Tillinghast Reflector Echelle Spectrograph (TRES)
on the FLWO 1.5m Tillinghast Reflector. TRES is a high-throughput
cross-dispersed fiber-fed echelle spectrograph. We used the medium
fiber ($2\farcs3$ diameter) for a resolving power of $R \simeq
44\,000$. The spectral resolution of the instrumental profile is 6.7
km s$^{-1}$ at the center of all echelle orders. For calibration
purposes, we acquired a thorium-argon lamp spectrum through the
science fiber both before and after every science spectrum. Exposure
times were 900s in good conditions, achieving a signal-to-noise ratio
of 15 per pixel at $7150\ {\rm \AA}$ (the pixel scale at this
wavelength is $0.059\ {\rm \AA ~pix^{-1}}$). These exposure times were
increased where necessary in poor conditions. The spectra were
extracted and processed using the pipeline described in
\citet{Buchhave(2010)}.

\section{Radial Velocities \& Orbit Determination} \label{sec:rv_analysis}

We derived radial velocities using standard cross-correlation
procedures based on the methods of \citet{Kurtz(1998)}. We used an
observed template spectrum of Barnard's Star, a slowly rotating
\citep[130.4 days,][]{Benedict(1998)} M4.0 dwarf
\citep{Kirkpatrick(1991)}, that was obtained on UT 2011 April 15.  We
performed correlations using a wavelength range of 7065 to 7165\AA ~in
order 41 of the spectrum, a region dominated by strong molecular
features due to TiO in mid-type M stars \citep{Irwin(2011b)}.

We adopt a Barycentric radial velocity of $-110.3 \pm 0.5\ {\rm
  km\ s^{-1}}$ for Barnard's Star, derived from presently unpublished
CfA Digital Speedometer \citep{Latham(2002)} measurements spanning 17
years. Barnard's Star and LHS~1610A both have negligible rotational
broadening at the resolution of the TRES spectra, so it was not
necessary to apply any rotational broadening to the template spectrum
prior to correlation.  The radial velocities derived from this
analysis are reported in Table \ref{tab:rv-data}.


\begin{deluxetable}{lrr}
\tabletypesize{\normalsize}
\tablecaption{Radial velocities of LHS~1610A \label{tab:rv-data}}
\tablecolumns{3}
\tablenum{2}

\tablehead{
\colhead{BJD\tablenotemark{a}} & \colhead{$v_{\rm
    rad}$\tablenotemark{b}\tablenotemark{c}} &
\colhead{$h$\tablenotemark{d}} \\ \colhead{(days)} & \colhead{(${\rm
    km\ s^{-1}}$)} &
}

\startdata
2457785.7131 & 28.448 & 0.941 \\
2457786.7850 & 32.365 & 0.940 \\
2457787.6378 & 35.502 & 0.943 \\
2457794.6483 & 22.514 & 0.948 \\
2457795.7182 & 26.224 & 0.945 \\
2457800.7416 & 44.533 & 0.935 \\
2457806.6698 & 27.585 & 0.936 \\
2457807.6875 & 31.293 & 0.903 \\
2457808.6590 & 34.944 & 0.931 \\
2457821.6194 & 43.586 & 0.933 \\
2457822.6458 & 45.893 & 0.906 \\
2457823.6552 & 40.479 & 0.860 \\
2457824.6210 & 25.451 & 0.915 \\
\enddata


\tablenotetext{a}{Barycentric Julian Date of mid-exposure, in the TDB
  time-system.}  
\tablenotetext{b}{Barycentric radial velocity.}
\tablenotetext{c}{Internal model-dependent uncertainties on each velocity are
  $\sigma$/$h$, where $\sigma$ is listed in Table \ref{tab:orb_el} and
  $h$ is the peak-normalized cross-correlation for each spectrum listed
  here.}
\tablenotetext{d}{Peak normalized cross-correlation.}
\end{deluxetable}

The useful radial velocity information content of the TRES spectra
gathered in our program for mid-M stars is dominated by the features
in order 41.  We find the velocities in the other orders have higher
scatter, and including them does not improve the results
significantly.  It is therefore not appropriate to use the rms of the
velocities in the individual orders to estimate the uncertainties in
our adopted order, as this would result in an overestimate.  Instead,
we derive the radial velocity uncertainties during fitting (e.g.,
\citealt{Gregory(2005)}). These internal model-dependent uncertainties
are $\sigma$/$h$, where $\sigma$ is the parameter from the MCMC
analysis found in Table \ref{tab:orb_el} and $h$ is the
cross-correlation, evaluated at the best-fitting velocity and
normalized to a peak value of one, as defined in \citet{Tonry(1979)},
for each spectrum listed in Table \ref{tab:rv-data}. Total
uncertainties on each absolute measurement should include the systemic
error of 0.5 km s$^{-1}$ from the Barnard's Star template.

The cross-correlation functions (CCFs) we created using the TiO
features in order 41, have a number of sidelobes surrounding the
central peak. We find a pair of prominent local maxima at
approximately $\pm 50\ {\rm km\ s^{-1}}$ from the central peak and
numerous other features at larger velocities.  These arise as a result
of the structure of the molecular bandhead, with lines being close to
evenly spaced in velocity.  This does not affect the radial velocities
determined from the cross-correlation peak, provided care is taken to
fit only the central peak, but presents some difficulty for detection
of additional stellar lines (e.g., due to additional components in
multiple systems) and other conventional analysis of the correlation
function such as line bisectors.


To alleviate this problem, we also perform a least-squares
deconvolution (LSD) of the target star spectrum against the observed
template spectrum (e.g., \citealt{Donati(1997)}).  Deconvolution is
prone to amplifying noise and producing spurious features,
particularly in the present case where the template has the same
resolution as the target. The target star spectra also tend to have
low signal-to-noise ratios (approximately 15, as noted above in \S
\ref{sec:data}), so we apply Tikhonov regularization
\citep{Tikhonov(1998)} and use features from several additional
surrounding orders in the red part of the spectrum in this analysis to
help with averaging out the noise.

\begin{figure}
\includegraphics[scale=.47,angle=0]{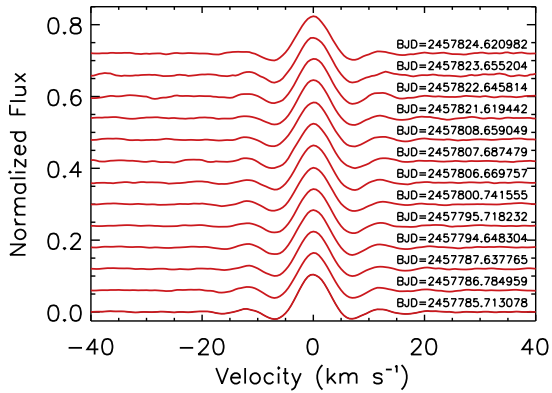}
\caption{Least-Squares Deconvolution (LSD) curves for each spectrum,
  shifted to a velocity of zero and stacked for clarity. Noted is the
  Barycentric Julian Date for each observation. There is no evidence
  of a second spectrum due to a stellar companion in any of the LSDs,
  nor is there any rotational broadening. \label{fig:lsd}}
\end{figure}

We show the LSD curves for the individual epochs in Figure
\ref{fig:lsd}.  As expected, these are compatible with
$\delta$-functions and show no indication of a second stellar spectrum
due to a companion, nor any additional rotational broadening in
LHS~1610A compared to Barnard's Star.

Having confirmed that the target is single-lined, we proceed to fit a
standard eccentric Keplerian orbit to the velocities derived from the
cross-correlation analysis using the {\sc emcee} package
\citep{Foreman-Mackey(2013)} to implement a Markov Chain Monte Carlo
(MCMC) sampler.  This model has 7 free parameters: the orbital period
$P$, epoch of inferior conjunction $T_0$, $e \cos \omega$, $e \sin
\omega$ (where $e$ is eccentricity and $\omega$ is argument of
periastron), systemic radial velocity $\gamma$, semiamplitude $K$ and
velocity uncertainty $\sigma$.  We use $e \cos \omega$ and $e \sin
\omega$ as jump parameters for mathematical convenience, but adopt
uniform priors in $e$ and $\omega$.  A modified Jeffreys prior of the
form $1 / (\sigma + \sigma_a)$ was used for $\sigma$ with $\sigma_a$
set to 10\% of the final value determined for $\sigma$.  In addition,
we use the estimated primary mass as a jump parameter in order to
propagate the uncertainty on the primary mass. We use a Gaussian prior
on the primary mass with the mean and standard deviation fixed to the
values reported in Table \ref{tab:system_info}. Uniform improper
priors were used for all other parameters.  The individual data points
were weighted by $h^2$ during fitting to account for the degradation
of the velocity precision in epochs with lower peak correlation.

\begin{figure}
\centering
\includegraphics[scale=.40,angle=0]{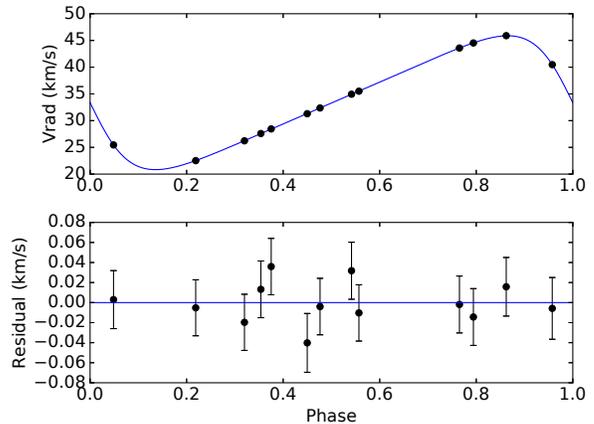}
\caption{Spectroscopic orbit of LHS~1610A from TRES (upper panel) and
  residuals (lower panel) after subtracting the best-fitting model,
  plotted as a function of normalized orbital phase (measuring from 0
  at inferior conjunction).  The average estimated velocity
  uncertainty is 28 m s$^{-1}$. \label{fig:orbit}}
\end{figure}

We ran simulations using $100$ chains initialized using a
Levenberg-Marquardt fit perturbed by $3 \sigma$ using independent
Gaussian deviates in each parameter. We ran chains for $6 \times 10^4$
samples, discarding the first $1 \times 10^4$ as a burn-in phase,
resulting in a combined total of $5 \times 10^6$ samples from the
posterior probability density function.  We report the resulting
parameters and uncertainties in Table \ref{tab:orb_el} using the
median and $68.3$ percentile of the absolute deviation of the samples
from the median as the central value and uncertainty, respectively. We
show the orbit in Figure \ref{fig:orbit}.

\begin{deluxetable}{lc}
\tablecaption{Orbital Elements for LHS~1610A \label{tab:orb_el}}
\tablecolumns{2}
\tablenum{3}
\tablewidth{0pt}
\tablehead{\colhead{Parameter}    &
           \colhead{Value}                          
}

\startdata
MCMC parameters\\
\hline
$e \cos \omega$                      & $0.00245 \pm 0.00148$ \\
$e \sin \omega$                      & $0.36941 \pm 0.00093$ \\
$T_0$ (BJD)                          & $2457781.739 \pm 0.011$ \\
$P$ (days)                           & $10.5918 \pm 0.0028$ \\
$\gamma$ (km s$^{-1}$)\tablenotemark{a} & $33.324 \pm 0.018$ \\
$K$ (km s$^{-1}$)                     & $12.527 \pm 0.017$ \\
$\sigma$ (km s$^{-1}$)                & $0.0265 \pm 0.0072$ \\
\hline
Derived parameters\\
\hline
$e$                                  & $0.36942 \pm 0.00093$ \\
$\omega$ (deg)                       & $89.62 \pm 0.23$ \\
$T_{\rm peri}$ (BJD)                   & $2457781.734 \pm 0.013$ \\
$a_1 \sin i$ (AU)                    & $0.011333 \pm 0.000016$ \\
$f_1(M)$ (${\rm M}_\odot$)            & $0.0017311 \pm 0.0000070$ \\
$q_{\rm min}$                          & $0.252 \pm 0.011$ \\
$a_{\rm min}$ (AU)                     & $0.0563 \pm 0.0020$ \\
$M_{2,{\rm min}}$ (${\rm M}_\odot$)      & $0.0428 \pm 0.0031$ \\
$M_{2,{\rm min}}$ (M$_{\rm Jup}$)         & $44.8 \pm 3.2$ \\
\enddata


\tablenotetext{a}{The uncertainty on the systemic velocity $\gamma$
  does not include the systematic uncertainty of 0.5 km s$^{-1}$ from
  the Barnard's Star template radial velocity.}


\end{deluxetable}

\section{MEarth Photometric Rotation Period} \label{sec:rot_per}

\begin{figure}
\centering
\includegraphics[scale=.45,angle=0]{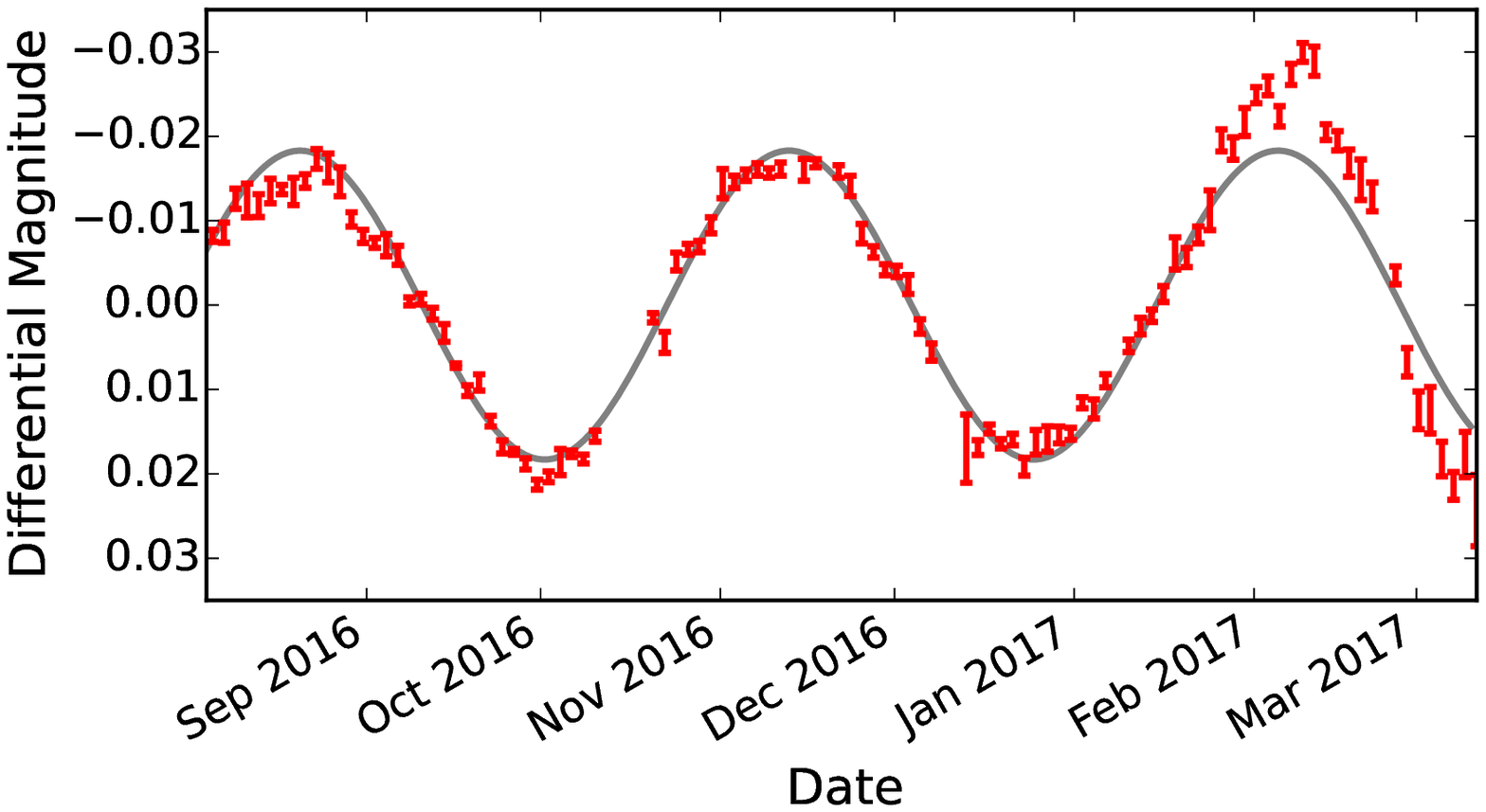}
\includegraphics[scale=.45,angle=0]{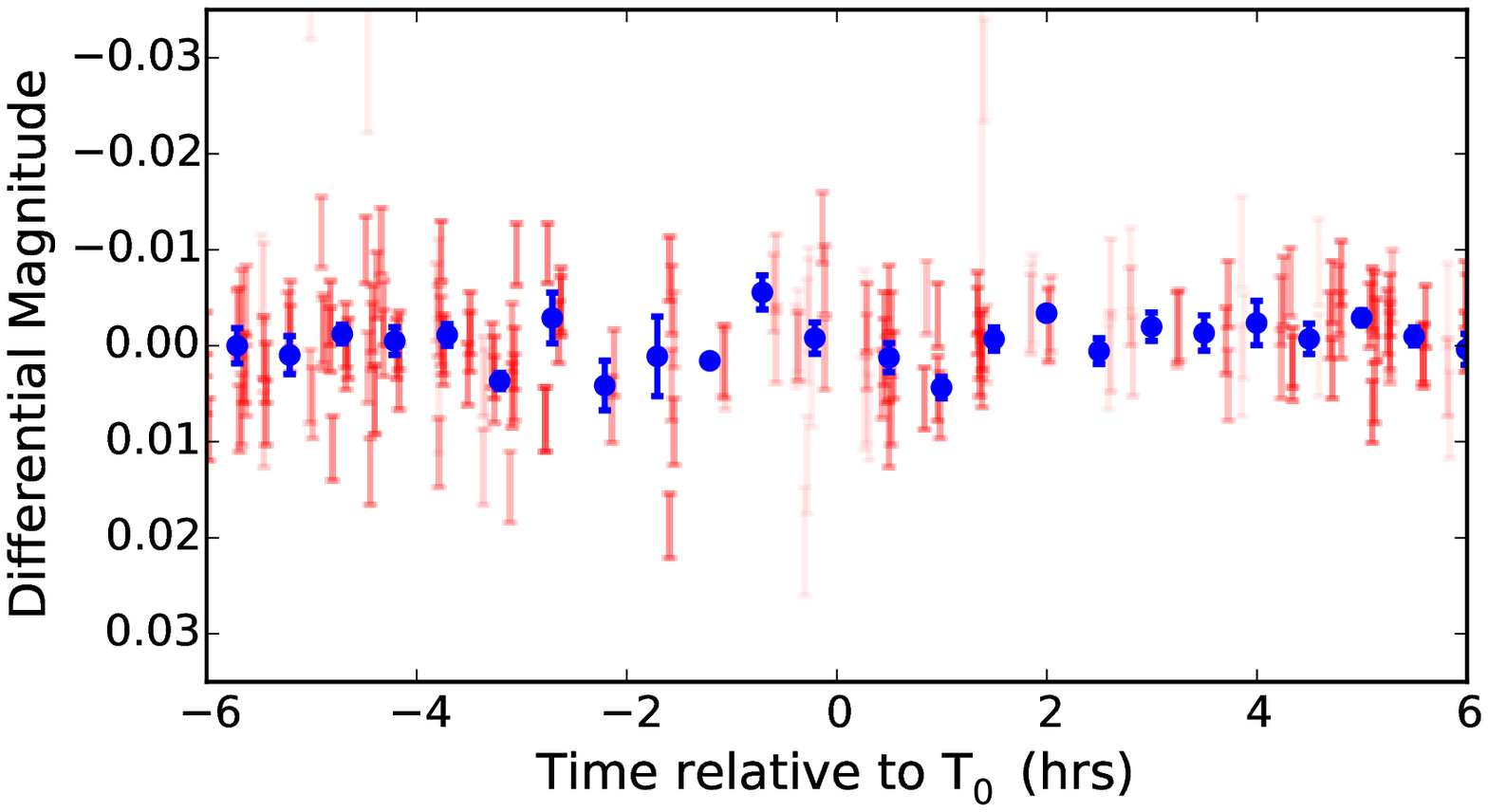}
\caption{{\it (Top)} Light curve for LHS 1610A using data from
  MEarth-South, binned by 2 days. The gray line shows a model
  sinusoid, with a rotation period of 84.3 days and a variability
  semi-amplitude of 0.018 mags. {\it (Bottom)} Phase-folded residuals
  from the light curve (top panel), after removing a 2-day running
  median. Red points are the unbinned data from the top panel. The
  opacity of the points indicates the size of the error, with larger
  error points being more transparent. The blue points are the median
  of half-hour intervals. The error bars on the blue circles are the
  standard error on the mean, using $1.48\times$ the median absolute
  deviation in place of the standard deviation. No eclipses are
  seen. \label{fig:rot_per_eclipse}}
\end{figure}

As part of the characterization of this system, we present here a new
photometric rotation period, measured using data from MEarth
\citep{Nutzman(2008),Irwin(2015)}. MEarth consists of eight robotic
telescopes located atop Mt. Hopkins in Arizona (MEarth-North), and
eight additional telescopes at the Cerro Tololo Inter-American
Observatory (CTIO) in Chile (MEarth-South). To improve the
determination of photometric rotation periods initially detected from
MEarth-North, LHS~1610 was re-observed from MEarth-South to take
advantage of the superior weather conditions at CTIO during the
appropriate observing season for this object.  We obtained data
spanning a full observing season on $172$ nights from UT 2016 August 4
to 2017 March 10 using a single telescope of the MEarth-South
array. We acquired 3697 exposures of 15s in groups of three
back-to-back exposures, with these groups or ``visits'' to the target
separated by approximately 30 minutes. Following our standard
differential photometry procedures, we reduced the data to light
curves, which we then analyzed as described in \citet{Irwin(2011)} and
\citet{Newton(2016)}. We show the resulting light curve in the top
panel of Figure \ref{fig:rot_per_eclipse}; the light curve data are
listed in an electronic-only table.

From this analysis, we determine a rotation period of 84.3 days with a
semi-amplitude of variability of 0.018 magnitudes. A small evolution
of the morphology of the modulation is seen toward the end of the
observing season. Our new period is consistent with our previous
detection of an 83.7 day rotation period using MEarth-North data, as
reported in \citet{Newton(2016)} and which was an update of the 78.8
day period reported in \citet{Irwin(2011)}. However, the new light
curve contains denser sampling over two complete rotation cycles and
is an improvement over our previous measurements. We assign this
object a ``grade A'' rotation period on the scale defined in
\citet{Newton(2016)}.


The phase coverage of the photometry is also sufficient to search for
eclipses. We look for a primary eclipse using the light curve from
MEarth-South, shown in the top panel of Figure
\ref{fig:rot_per_eclipse}. To remove the stellar variability, which is
not quite sinusoidal, we apply a running 2-day median filter. We then
phase-folded the data using the period and ephemeris listed in Table
\ref{tab:orb_el}. After rejecting outliers with absolute relative flux
greater than 0.02 mag, we find the median absolute deviation is 0.0035
and 0.0017 for unbinned and binned data, respectively. As is shown in
the bottom panel of Figure \ref{fig:rot_per_eclipse}, no eclipses are
present in the data.


\section{Constraints on the companion} \label{sec:constraints}

\subsection{Photometry and Astrometry} \label{subsec:phot}

Because both good quality $VRIJHK$ photometry and accurate
trigonometric parallaxes exist for this object, we compare the
photometric distance estimate, calculated using the distance relation
in \citet{Henry(2004)}, with the trigonometric distance to place upper
limits on the mass of the secondary component. An equal-luminosity
companion would result in the overluminosity of the system and its
photometric distance estimate would therefore be underestimated by a
factor of $\sqrt{2}$ when compared to its trigonometric distance. We
find a photometric distance estimate of 9.7 $\pm$ 1.5 pc, in agreement
with the trigonometric distance of 9.9 $\pm$ 0.2 pc. We can therefore
infer that the companion is not of equal luminosity.




Companions with magnitude differences ($\Delta$mag) of 2.5 (flux
ratios $=$ 10) from their primaries are reliably detected using the
{\sc todcor} package \citep{Zucker(1994)}. The TiO-bands that we use
for analysis are effectively in the $I-$band, so we compare the $M_I$
of LHS~1610A to that of two known late M dwarfs: 2MASS~J2306-0502
(also known as TRAPPIST-1), an M7.5 V \citep{Cruz(2003)} and
SCR~J1845-6357A, an M8.5 V \citep{Henry(2006)}. We note that
SCR~1845-6357A is known to have a T dwarf companion, but this companion
contributes a negligible amount of light in the $I-$band. The $M_I$ of
LHS~1610A, 2MASS~J2306-0502, and SCR~J1845-6357 are 10.68, 13.60, and
14.45 mag, respectively. The magnitude differences in $I$ between
LHS~1610A and the two late M dwarfs are larger than 2.5 mag, so it is
not likely that we would have detected a companion of spectral type
M7.5 V or M8.5 V in our optical spectra. However, the $M_K$ of the
three stars are 8.08, 9.79, and 10.50 mag, respectively, resulting in
$\Delta$mag $<$ 2.5, so it is possible that we could detect an M7.5
and M8.5 dwarf in infrared spectra; see \S \ref{subsec:ir_spec}.


%

\subsection{Age Estimate} \label{subsec:age}

In order to use evolutionary models to estimate an upper limit on the
mass of the companion, we require an estimate of the age of the
system. Our systemic velocity determination permits the calculation of
galactic space motions, relative to the local standard of rest, using
the method outlined in \citet{Johnson(1987)}. We find velocities of
-30.5 $\pm$ 0.5, -32.0 $\pm$ 0.8, and -21.3 $\pm$ 0.4 km s$^{-1}$ for
$U_{\odot},V_{\odot},W_{\odot}$, respectively, where $U_{\odot}$ is the
radial component, positive in the direction of the Galactic center,
$V_{\odot}$ is the azimuthal component, and $W_{\odot}$ is the
vertical component. Using these space velocities and the method
described in \citet{Bensby(2003)}, we calculate a probability of only
1\% that the object belongs to the thick disk population, as opposed
to the thin disk population. We therefore deem LHS~1610A a member of
the thin disk population, to which \citet{Bensby(2003)} assign an
average age of 4.9 $\pm$ 2.8 Gyr. Using the rotation period-age
relation from \citet{Newton(2016)}, we also conclude, due to its long
rotation period, that the system is likely at least 4.5 Gyr old. We
note that because the two age estimates agree, the rotation period of
the primary has not been affected by the presence and close proximity
of the secondary.

With this age estimate, we perform a linear interpolation of the 1 and
5 Gyr COND03 evolutionary models \citep{Baraffe(2003)} to estimate an
upper limit on the mass and effective temperature of an object with an
$M_K$ of 10.50 mag (i.e., SCR~J1845-6357A, as described above). For an
object with an age of 1 Gyr, this results in a maximum mass and
effective temperature of 0.082 M$_{\odot}$ and 2436 K; for a 5 Gyr-old
object, we calculate a maximum mass and effective temperature of 0.084
M$_{\odot}$ and 2444 K.




%
%
%
%

\vspace{1cm}
\subsection{Infrared Spectroscopy} \label{subsec:ir_spec}



 
%


To place further constraints on the secondary component, we observed
LHS~1610A using the Immersion GRating INfrared Spectrometer
\citep[IGRINS,][]{Yuk(2010)} on the 4.3-meter Discovery Channel
Telescope (DCT) in Happy Jack, Arizona, on the nights of UT 2017
September 25 and 26. IGRINS is a cross-dispersed, high-resolution (R =
$\lambda/\Delta\lambda$ = 45,000) near-infrared spectrograph with a
wavelength coverage of 1.45 to 2.5 $\mu m$, which obtains simultaneous
observations in both the $H$ and $K$ bands
\citep[][]{Yuk(2010),Park(2014),Mace(2016a),Mace(2016b)}. We
calculated exposure times to achieve a signal-to-noise ratio of
approximately $150$ per wavelength bin. We observed the A0~V telluric
standard stars HR~8422 and HR~945 either immediately before or after
and within 0.1 airmasses of LHS~1610A. We used the publicly available
reduction pipeline for IGRINS \citep{plp} to process the spectra and
{\sc xtellcor\_general} \citep{Vacca(2003)} to remove the telluric
lines.

To measure the IGRINS radial velocities, we followed the method
described in \cite{Han(2017)}. We used the ephemeris from the TRES
spectroscopic orbit to determine that the orbital phases for the
system were 0.67 (night one) and 0.77 (night two), near the maximum
radial velocity separation. We did not detect the signal of the
secondary component in the IGRINS data.

To place an upper limit on the mass of the secondary component, we
injected BT-Settl models \citep[][]{Allard(2012)} of objects with
effective temperatures ranging from 2000 K to 3100 K (cool brown
dwarfs to spectral type M4), into the IGRINS spectra. We injected each
BT-Settl model at different RV shifts of the primary component,
calculated based on a grid of masses that ranged from 0.17 M$_\odot$,
corresponding to an equal mass companion, to 0.042 M$_\odot$, the
lower mass limit determined by the TRES orbital solution. Before
injection, we matched the resolution of BT-Settl models to that of the
IGRINS data and added photon noise corresponding to the expected
brightness of the putative secondary. We assumed the radii of the
primary and secondary components to be 0.15 R$_{\odot}$ and 0.10
R$_{\odot}$, respectively. After injecting the secondary signal into
the IGRINS spectrum, we cross-correlated the simulated LHS~1610A
spectrum with the BT-Settl synthetic spectra and searched for the mass
where the companion became undetectable.

\begin{figure}
\centering
\includegraphics[scale=.55,angle=0]{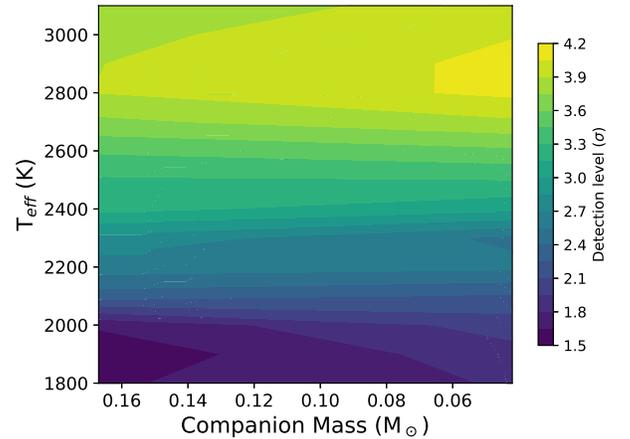}
\caption{Illustrated is effective temperature versus mass for
  LHS~1610B, with darker colors indicating a less significant
  detection. The results of our injection and recovery analysis
  indicate that we would have been able to detect to 3$\sigma$ a
  companion at any mass with an effective temperature of roughly 2500
  K, indicating that the companion is cooler than this temperature.
  \label{fig:igrins}}
\end{figure}

We show the results of our injection and recovery analysis in Figure
\ref{fig:igrins}. Plotted is the effective temperature of the BT-Settl
models that we used versus the simulated secondary masses. The color
bar indicates the detection level, which corresponds to the height of
the cross-correlation peak in terms of the standard deviation of the
entire CCF. It is evident that the effective temperature has a larger
effect on the detection than the mass. We place an upper limit of 2500
K to 3$\sigma$ on the effective temperature of the companion that we
could have detected with our IGRINS data. We therefore conclude that
the companion is not likely to be an M dwarf.

\section{Discussion} \label{sec:discussion}

LHS~1610 was noted as a double-lined spectroscopic binary in
\citet{Bonfils(2013)}. However, inspection of our initial TRES
spectrum of this object did not reveal the second line indicative of a
nearly-equal luminosity stellar companion. We inspected the publicly
available HARPS-GTO spectra of this object to determine whether our
non-detection was due to the lower resolution of TRES, compared to
that of HARPS, but did not see a second set of lines in those
data. This object was not included in the sample of
\citet{Tokovinin(1992b)}, a work that searched for brown dwarf
companions to M dwarfs, as the cooler spectral type limit of the
sample was M3 V. As noted in Table \ref{tab:system_info}, LHS~1610A
has a spectral type of M4.0 V \citep{Henry(2006)}.

Preliminary work from \citet{Udry(2000)} showed for a small sample of
M dwarf binaries that most systems with orbital periods of less than
roughly 10 days have orbits that are nearly circular, similar to
results for solar-type binaries
\citep[e.g.,][]{Duquennoy(1991),Latham(2002)}. Thus, with a 10.6-day
orbit, the eccentricity of the system is not surprising, as its period
is not short enough to have circularized.


We list the systems consisting of an M dwarf and a brown dwarf within
10 pc in Table \ref{tab:other_mdbd}. Of note is the scarcity of such
systems: only four of the approximately 200 M dwarf systems within 10
pc \citep{Henry(2016)} are known to harbor a brown dwarf
companion. The primary component of GJ~229 is an early M dwarf, while
GJ~569B, WIS~J0720-0846A, and SCR~J1845-6357A are all late M
dwarfs. There are no reports of a mid-M dwarf within 10 pc in the
literature with a confirmed brown dwarf companion. We do, however,
note that there are two other nearby mid-M dwarfs suspected to have
brown dwarf companions that have yet to be confirmed: GJ~595A
\citep{Nidever(2002)} and GJ~867B \citep{Davison(2015)}.

The range of mass ratios (0.50 -- 0.25) for this system from the
upper and lower mass limits (0.084 -- 0.043 M$_{\odot}$) on the
companion places it in a region of distribution space that is
currently sparsely populated for M dwarfs \citep{Wintersphd}. We note
that it is possible that the companion is an early L dwarf that lies
above the hydrogen-burning limit of approximately 2075 K
\citep{Dieterich(2014)}. However, if we assume that the average value
of $\sin$$^3$$i$ is $3$$\pi$/$16$, then the mass is on average a
factor of 1.7 larger than the minimum mass. This would result in a
mass for the secondary of roughly 0.073 M$_{\odot}$, which is just at
the 0.070 -- 0.075 M$_{\odot}$ mass boundary between stars and brown
dwarfs \citep{Benedict(2016),Dupuy(2017)}.

\begin{deluxetable*}{lclccccccc}
\tabletypesize{\small}
\tablecaption{Information for 10 pc MD$+$BD Systems \label{tab:other_mdbd}}
\tablecolumns{10}
\tablenum{4}
\tablewidth{0pt}
\tablehead{\colhead{Name}                &
	   \colhead{RA}                  &
           \colhead{DEC}                 &
	   \colhead{d}                   &
           \colhead{Ref}                 &
	   \colhead{a}                   &
           \colhead{SpT$_{\rm pri}$}          &
	   \colhead{SpT$_{\rm sec}$}          &
           \colhead{Ref}                 &
           \colhead{M$_{\rm pri}$}            \\
	   \colhead{   }                 &
           \colhead{(hh:mm:ss)}          &
           \colhead{(dd:mm:ss)}          &
	   \colhead{(pc)}                &
           \colhead{     }               &
           \colhead{(au)}                &
	   \colhead{   }                 &
           \colhead{   }                 &
           \colhead{   }                 &
           \colhead{(M$_{\odot}$)}           
}

\startdata
GJ~229AB                 & 06:10:34.6  & $-$21:51:52  &  5.8 $\pm$ 0.01  & 10,9,6  &  58    &  M1.5 V   &  T6        &   4  & 0.60 $\pm$ 0.02 \\  
WIS~J0720-0846AB         & 07:20:03.3  & $-$08:46:50  &  7.0 $\pm$ 0.02  & 8       &   1.2  &  M9.5 V   &  T5        &   2  & 0.075 $\pm$ 0.02 \\  
GJ~569BC                 & 14:54:29.2  & $+$16:06:03  &  9.9 $\pm$ 0.13  & 10,9    &   0.9  &  M8.5 V   &  M9        &   5  & 0.076 $^{+0.0086}_{-0.0076}$\tablenotemark{*} \\  
SCR~J1845-6357AB         & 18:45:02.0  & $-$63:57:47  &  3.9 $\pm$ 0.02  & 3,7     &   5.7  &  M8.5 V   &  T6        &   1  & 0.075 $\pm$ 0.02 \\  
LHS~1610AB               & 03:52:41.0  & $+$17:01:04  &  9.9 $\pm$ 0.2   & 10,7    &   0.06 &  M4.0 V   &  \nodata   &   7  & 0.17 $\pm$ 0.02 \\  
\enddata


\tablecomments{Listed are the name of the pair, the epoch 2000.0
  coordinates, the distance and error in parsecs, followed by the
  reference(s) for the parallax(es) from which the distance was
  calculated, and the projected linear separation in AU. In the cases
  of GJ~229AB, WIS~J0720-0846AB, and SCR~J1845-6357AB, where no orbits
  exist, the calculated projected linear separation (angular
  separation $\times$ distance) has been multiplied by a factor of
  1.26, as prescribed in \citet{Fischer(1992)}. Next are listed the
  spectral types of the primary and secondary, along with the
  reference. Finally, the masses of the primary components are
  listed. These masses have been estimated using the
  $M_K$~mass-luminosity relation from \citet{Benedict(2016)}.}

\tablenotetext{*}{GJ~569BC is the only pair with measured dynamical
  masses, for which \citet{Dupuy(2017)} report M$_{\rm pri} =$ 80
  $^{+9}_{-8}$ M$_{\rm Jup}$ and M$_{\rm sec} =$ 58 $^{+7}_{-9}$ M$_{\rm Jup}$.}

\tablerefs{ 
(1) \citet{Biller(2006)};
(2) \citet{Burgasser(2015)};
(3) \citet{Deacon(2005b)}; 
(4) \citet{Dieterich(2012)}; 
(5) \citet{Dupuy(2017)}; 
(6) \citet{Gaia(2016)};
(7) \citet{Henry(2006)};
(8) \citet{Scholz(2014)};
(9) \citet{vanLeeuwen(2007)};
(10) \citet{vanAltena(1995)}.
}
\end{deluxetable*}



Future work will enable a better constraint on the secondary component
of this system. For example, because the companion is unequal in both
flux and mass, this system should exhibit an astrometric perturbation
on the photocenter of the system \citep{vandeKamp(1975)}. The
magnitude of the perturbation, which we estimate to be approximately
7.5 mas, should be detectable by {\it Gaia}. An astrometric orbit from
{\it Gaia} will provide the inclination for the system and permit the
calculation of dynamical masses for the two components. Our TRES
spectroscopic orbit will provide the necessary ephemeris for the
astrometric orbital solution.




\acknowledgments

The authors thank the anonymous referee both for their prompt response
and for their comments and suggestions. We thank Samuel Quinn,
Guillermo Torres, Sergio Dieterich, Trent Dupuy, Wei-Chun Jao, Jayne
Birkby, and Matthew Payne for illuminating discussions and
suggestions. We especially thank Todd Henry for his assistance with
the list of nearby M dwarf - brown dwarf binaries. We express our
gratitude to Jessica Mink for the processing and extraction of the
TRES spectra.

The MEarth Team gratefully acknowledges funding from the David and
Lucille Packard Fellowship for Science and Engineering (awarded to
D.C.). This material is based upon work supported by the National
Science Foundation under grants AST-0807690, AST-1109468, AST-1004488
(Alan T. Waterman Award), and AST-1616624. E.R.N. is supported by an
NSF Astronomy and Astrophysics Postdoctoral Fellowship. This
publication was made possible through the support of a grant from the
John Templeton Foundation. The opinions expressed in this publication
are those of the authors and do not necessarily reflect the views of
the John Templeton Foundation. 

P.S.M. and E.H. acknowledge support from the NASA Exoplanet Research
Program (XRP) under Grant No. NNX15AG08G issued through the Science
Mission Directorate.

These results made use of the Discovery Channel Telescope at Lowell
Observatory, supported by Discovery Communications, Inc., Boston
University, the University of Maryland, the University of Toledo and
Northern Arizona University. This work used the Immersion Grating
Infrared Spectrograph (IGRINS) that was developed under a
collaboration between the University of Texas at Austin and the Korea
Astronomy and Space Science Institute (KASI) with the financial
support of the US National Science Foundation under grant AST-1229522,
of the University of Texas at Austin, and of the Korean GMT Project of
KASI.

%

\vspace{5mm}
\facilities{FLWO 1.5m (TRES)}, {MEarth}, {DCT 4.3m (IGRINS)}


\bibliographystyle{aasjournal}
\bibliography{/home/jwinters/references/masterref}

\clearpage


\end{document}